\documentstyle[12pt]{article}

\def\be{\begin{equation}}
\def\ee{\end{equation}}

\newcommand{\eq}[1]{(\ref{#1})}
\def\ln{{\rm ln}}
\begin{document}
\title{
Grand unification with gauge mediated supersymmetry breaking%
\footnote{Talk presented by S.T. at the 10th International Seminar
``Quarks-98'', Suzdal, Russia, May 18-24, 1998; to appear in the Proceedings.}
}
\author{ S.L.Dubovsky, D.S.Gorbunov, and S.V.Troitsky\\
{\small{\em
Institute for Nuclear Research of the Russian Academy of Sciences
}}
}
\date{}

\maketitle

\begin{abstract}
\small
We consider constraints on gauge mediated supersymmetry breaking models
imposed by the requirement of grand unification. 
In particular, we demonstrate several ways to reduce
the number of parameters coming from the dynamical supersymmetry
breaking sector. One of approaches exploits nonperturbative
unification of gauge couplings in multi-messenger models.
\end{abstract}

{\bf 1.} {\sl Introduction.}  Almost all
existing experimental particle physics data fit well the Standard
Model predicitons, so every reason to invent new physics beyond this
theory is purely aesthetical. From the particle theory point of view,
more beautiful models are obviously those which have more symmetries,
and those which are predictive, i.e. have less free parameters. Two
of the most popular extensions of the Standard Model deal with supersymmetry
and with Grand Unification, respectively. The latter concept shares
both pleasant features, namely, larger symmetries in gauge interactions
and in matter content together with restrictions on the existing
parameters of the Standard Model. On the other hand, supersymmetry
requires much more symmetries restricting particle dynamics but often
fails to be predictive in realistic models --- it introduces quite a few new parameters
instead of constraining the existing ones. Merging the two concepts,
however, can lead to really aesthetically appealing models --
Supersymmetric Grand Unified Theories (GUTs).

The main purpose of this talk is to demonstrate how the concept of Grand
Unification helps to constrain the parameters in one particular class
of supersymmetric models -- models with gauge mediated supersymmetry
breaking (GMSB). The latter models are known to be rather predictive
by themselves; we show here that in some cases the number of free
parameters may be even more reduced once a particular model is
specified. We begin with a brief sketch of GMSB models, then discuss
possible general constraints coming from the Grand Unification
postulates. We turn then to particular examples and consider the ways
the 
models of direct gauge mediation may be constrained.

{\bf 2.} {\sl Sketch of GMSB.} 
In most realistic supersymmetric models,
supersymmetry is broken in the sector of fields different from the
Standard Model and its minimal supersymmetric extension (MSSM). The
(rather weak) interaction between the two sectors transfers
supersymmetry breaking to the visible fields, and so
determines the parameters of the MSSM. In the most general case, there
are of order $10^2$ of these parameters, and to start comparing the model with
experiments, one first needs to constrain the parameter space. The most
attractive way is to specify the interaction between the two sectors
and to calculate the values of the parameters, but in practice several
relations between them 
are often conjectured.

It is the kind of interaction between the dynamical supersymmetry
breaking (DSB) sector and the observable world which distinguishes between
different classes of supersymmetric models. This interaction should be
suppressed at low energies, and two conventional choices are
gravitational (which operates at energies of order Planck scale) and
Yukawa or/and gauge (operating somewhere between the electroweak and
Planck scales) interactions. The latter case corresponds to the GMSB
theories, and makes it possible to calculate the
supersymmetry breaking parameters of the MSSM by means of the field
theoretical methods, without invoking less understood theories
of quantum gravity. In the former case, some simple
constraints on the MSSM parameters are often postulated (see, e.g., Ref.\
\cite{Nilles} for  a review of gravity mediated supersymmetry
breaking). 

The key ingredient of the GMSB scenario is a set of messenger fields
which fall in the vectorlike multiplets of the Standard Model gauge
symmetry. They are the only fields from the visible sector which 
directly 
interact with the DSB sector. By means of this interaction (either
pure Yukawa in the minimal models or both Yukawa and gauge in the
so-called direct mediation models) the messengers obtain supersymmetry
breaking masses of their component fields. The parameters of the MSSM --
soft gaugino and scalar masses and trilinear couplings --- are generated
via loop effects by MSSM gauge interactions between messengers and
ordinary particles. Thus, the masses of the superpartners are determined by
their $SU(3)\times SU(2) \times U(1)$ quantum numbers and the
parameters describing the spectrum of messengers (typically, there are
two such parameters, the mass of the fermionic component, and the mass
splitting between bosonic messengers). The gauge mediation mechanism
naturally suppresses flavour violating processes and is highly
predictive --- all MSSM parameters (not counting 17 parameters of
the non-supersymmetric Standard Model) are calculated in terms of four
--- two in the messenger sector and two in the Higgs sector. The GMSB
phenomenology and model building are widely discussed, e.g.,
in the review Ref.\ \cite{Review}; here we
will concentrate on ways to further reduce the number of free
parameters by making use of Grand Unification constraints.

{\bf 3.} {\sl Simple GUT constraints.}

Let us list the well known
constraints which are characteristic to Grand Unified theories and
enable one to restrict parameters of the low-energy theory. These
constraints are 
\begin{enumerate}
\item
unification of gauge couplings;
\item
unification of Yukawa couplings; 
\item
matter content in full GUT multiplets (or explanation of splitting, as
in the case of Higgs doublets and triplets);
\item
interactions which may be written in terms of GUT multiplets.
\end{enumerate}
All these types of constraints may be used to rule out some models,
and the first and the second constraints may provide quantitative
bounds on parameters of a given model. Consider, as an example, the
minimal gauge mediated model with one (5+$\bar 5$) set of
messengers. Perturbative gauge unification in the visible sector is
unchanged by construction (at least in one loop). One may, however,
impose an additional constraint (which seems very plausible)
that the Grand Unified Theory, if exists, should contain the secluded
(DSB) sector as well. With this assumption, one can exploit the
unification of gauge couplings of the DSB and visible sectors to gain
information about the values of the coupling constant in the secluded
sector and, notably, about the scale where it
becomes strongly coupled. The latter scale $\Lambda_s$ determines the
value of one of the supersymmetry breaking parameters for a given model.
Phenomenological and cosmological constraints on this
parameter allow us to extract ``unifiable'' models from a plethora of
known DSB schemes. Only three of them (the ``3-2'' model \cite{ADS-85},
its extension with extra matter, and the model with $SU(2)$ group and
vector-like matter content \cite{Izawa}) were found to satisfy these
criteria \cite{Izawa,couplings} without introducing additional matter
thresholds.   Another example concerns the restrictions on the minimal
$SU(5)$ GUT with GMSB that come from the analysis of gauge and $b$--$\tau$
Yukawa coupling 
unification \cite{GUT_fall}. These restrictions in fact have shown the
inconsistency of the minimal model with unification.

One should note, however, that a unified theory containing both DSB
and visible sectors is still missing. Nevertheless, as we will see in the
next section, even the requirement of unification in the visible
sector may be sometimes very restrictive.

{\bf 4.} {\sl Constraining multi-messenger direct gauge mediation scenarios.}
Let us turn to the models with direct gauge mediation where the
messenger fields themselves are part of the DSB sector, namely, they
carry charges under the secluded gauge group. 
From the visible sector viewpoint, the latter gauge group plays a
role of a flavour group, so several copies of messenger fields should
appear in the spectrum -- the number of copies being equal to the
dimensionality of the corresponding representation of the secluded gauge
group. Since the messengers are charged under the Standard Model too,
their contribution to  the gauge beta functions of the Standard Model
may lead to the loss of the asymptotic freedom. For example, if the
number of $5+\bar 5$ multiplets exceeds four, the gauge coupling
constants become large below $M_{\rm GUT}\sim 10^{16}$~GeV which
contradicts the idea of perturbative unification. On the other hand,
it is often that the DSB group is not very small, so it has no
representation with dimension 4 or less\footnote{Recently, a model
of direct mediation with the DSB group $SU(2)$ and two sets of $5+\bar
5$ messengers has been constructed \cite{Agashi}. We will not discuss
this model here since analysis in this section  is valid for
multi-messenger models only.}. The most common approach to the question of
saving gauge unification in this case is to make the messengers
heavy
\cite{direct,Dimop-1}.
For high enough thresholds of the messengers, Landau poles might be
``pushed'' away to energies higher than $M_{\rm GUT}$. Though
particle phenomenology does not suffer from messengers being heavy,
they often are problematic for cosmology \cite{Dimop-1,new-nucl}. This
fact suggests to look for other ways of solving the gauge unification
problem in multi-messenger models. One of the ways is to invoke completely
new physics at intermediate scales between $M_{\rm SUSY}$ and $M_{\rm
GUT}$. If the Standard model fields and/or messengers are composite
states made of a few fundamental preons, the latter transforming as
complete GUT multiplets, then at high energies only preons contribute
to the beta functions of $SU(3)\times SU(2)\times U(1)$. Generally,
their contribution is smaller than the contribution of the composite
fields, and the coupling constants at $M_{\rm GUT}$ can remain
small. A few examples of gauge mediated supersymmetry breaking models
with direct gauge mediation and compositeness were constructed, but
they are either toy models \cite{we:compose} or too complicated to be
realistic \cite{compose}. All models require dynamical assumptions
about uncalculable dynamics at strong coupling.

Let us turn now to the last, and the least explored possibility. Its
main idea is to replace the perturbative unification of couplings by a
controllable and phenomenologically acceptable unification at the
strong coupling.  

The possibility of gauge coupling unification in the strong coupling
regime has been considered in the framework of both the Standard Model
and its supersymmetric extensions
\cite{Maiani}. Recently, this problem attracted
some interest again
\cite{nonpert-unif} after more precise measurements of the 
gauge coupling constants at $M_Z$ have been carried out. The latter results
differ from the two-loop unification predictions by more than one
standard deviation (see, e.g., Ref.\ \cite{kabak}). 

Note that running gauge couplings of the MSSM $\alpha_1$ and
$\alpha_2$ increase with energy, so $SU(2)\times U(1)$ is not
asymptotically free. These couplings, however, run relatively slow, so
Landau poles of these two groups appear at energies higher than the
unification scale. Together with the asymptotic freedom of QCD this
means that below $M_{GUT}$ all gauge couplings are small, and
perturbative analysis is valid. This picture impies the existence of
the ``desert'', i.e. absence of particles huge region of masses
between superparticle and unification scales. When new particles, like
several multiplets of messengers, are introduced, the first
coefficients of the $\beta$ functions increase, so gauge couplings may
become large at the unification scale.

Despite the fact that unification in this case occurs at the strong
coupling, it is unexpectedly controllable from the low energy point of
view, especially in the supersymmetric case \cite{Ross}. 
Consider one-loop evolution of the coupling constants in an asymptotically
non-free unified theory. If
$M_G$ is the unification scale and $\alpha_G$ is the value of the unified
gauge coupling at that scale, then the renormalisation group
equations 
\begin{equation}
{d\alpha_i\over dt}=b'_i\alpha_i^2
\label{RG}
\end{equation}
have a solution
$$
\alpha_i^{-1}(Q)=\alpha_G^{-1}+b'_i t,
$$
where $t={1\over 2\pi}\ln{Q\over M_G}$ and $b'_i$ 
are the first coefficients of the beta functions of the gauge
couplings in the asymptotically non-free theory.
Consider running of the {\em ratios} of pairs of the gauge couplings,
\begin{equation}
{d\over dt}\ln{\alpha_i\over\alpha_j}=b'_i\alpha_i-b'_j\alpha_j.
\label{r}
\end{equation}		
At one loop, these ratios have
infrared fixed points,
$$
{\alpha_i\over\alpha_j}={b'_j\over b'_i}.
$$
These fixed points are reached at the energies which are model-dependent
and may be read out from the solution to eq.\eq{r},
$$
{\alpha_i(Q)\over\alpha_j(Q)}={\alpha_G^{-1}+b'_j t \over
\alpha_G^{-1}+ b'_i t}.
$$
The condition that the fixed point is almost reached is
$|t|\gg\alpha_G^{-1}/b'_i$.  In the case of MSSM without additional
matter, one has
$\alpha_G^{-1}\sim 24$, so that for $\alpha_2$ the fixed point occurs
at $|t|\gg 24$, i.e., at $Q\ll
M_G\cdot\exp(-48\pi)\sim 10^{-66} M_G$ which certainly rules out the
possibility of the fixed point analysis.However, with new matter
added, the situation changes drastically --- $b_i$ increase and
$\alpha_G^{-1}$ decreases. Suppose that additional (messenger)
multiplets fall in the complete vector-like representations of the
$SU(5)$ unified gauge group, for example, ($5+\bar{5}$) or
($10+\bar{10}$).  
Then $b'_i=b_i+n$, where $b_i$ are $\beta$ function coefficients of
the MSSM.
Each ($5+\bar{5}$) set adds 1 to $n$ while
each ($10+\bar{10}$) adds 3. For $n\ge 5$ the unification occurs
at strong coupling. To estimate the energy scale where ratios of
couplings get close to the fixed point value, let us take
$\alpha_G=1$. Then even for $n=5$ the ratios are almost constant at
$Q<0.04 M_G$. 

For given $n$, the threshold corresponding to messenger mass
is uniquely determined. Indeed, the low energy running of MSSM couplings is
known, and at the threshold the couplings {\em should} have the ratios
equal to $b'_i/b'_j$. The energy where ratios of MSSM running
couplings, 
determined experimentally at $M_Z$, get to their fixed point values
corresponds to the messenger threshold. Note that at $n\ge 5$ the
corresponding thresholds are deep in the region of the attraction of the fixed
points. For $n=5$, for example, the threshold is between 1
and 10~TeV, much lower than $0.04\cdot M_G$. This means that the
fixed-point approach is self-consistent. The values of thresholds can
be read out from Ref.\cite{Ross}; values of $6\le n \le 20$ are
consistent with current bounds on the messenger mass \cite{to-appear}.

So, from the low energy MSSM point of view we just have new boundary
conditions for running of the gauge couplings. Instead of requiring
the equality of couplings at
$M_G$ (as in the case of perturbative unification), one should fix their
ratios at the messenger scale. Details of evolution of the couplings
near $M_G$, where they are large, are unknown; however, they do not affect
significantly the low-energy predictions
\cite{Ross,NewRoss}.

We conclude that having quite large number of messenger fields 
at scales between 
 $M_{\rm SUSY}$ and $M_{\rm GUT}$
does not contradict the gauge coupling unification, the latter
occuring in 
the nonperturbative regime. 

The most interesting feature of this scenario is that the strong unification
constrains significantly the parameter space of multi-messenger gauge
mediation models. Namely, the mass scale of the messenger fields --
one of the two parameters describing the superpartner masses -- is
determined for a given effective number of messengers $n$. 

It is worth noting, however, that the superparticle spectrum does not
depend significantly on the messenger mass $M$. Instead, the scale of
superpartner masses is set by the product $\Lambda=Mx$, where $0<x<1$.
In most supersymmetry breaking models, the scale $\Lambda$ depends on
the scale $\Lambda_S$ where the gauge coupling constant of the
secluded sector becomes strong. To determine $\Lambda_S$, one has to
put some boundary conditions on that gauge coupling. This can be done
either at $M_{\rm GUT}$, from the condition of ``total'' unification,
like in Ref.\cite{couplings}, or at some intermediate scale by means
of the fixed point formalism. If all contributions to the soft $B_\mu$
term come from the gauge mediation, and the supersymmetric $\mu$
parameter is tuned so that electroweak symmetry breaking is radiative,
then $\Lambda$ and $M$ determine the superparticle spectrum completely
--- the model thus has no free parameters. Whether this spectrum is
realistic, depends on the choice of model \cite{to-appear}.

{\bf 5.} {\sl Conclusion.}  
We have demonstrated that simple requirements
of the Grand Unification of both visible and secluded sectors, and/or
(in the multi-messenger case) of non-perturbative Grand
Unification in the visible sector can significantly reduce the
number of parameters of theories with gauge mediated supersymmetry
breaking, thus ruling out some of them or predicting
superparticle spectrum for others. 

We thank M.V.Libanov, Yu.F.Pirogov,
and V.A.Rubakov for interesting discussions. This work is supported in
part by the RFFI grant 96-02-17449a.
Work of S.T. is supported in part
by the U.S.\ Civilian Research and Development Foundation for
Independent States of FSU (CRDF) Award No.~RP1-187.
\def\prl#1#2#3{{\it Phys. Rev. Lett. }{\bf #1~} #3 (19#2)}
\def\np#1#2#3{{\it Nucl. Phys. }{\bf B#1~} #3 (19#2)}
\def\pr#1#2#3{{\it Phys. Rev. }{\bf D#1~} #3 (19#2)}
\def\pl#1#2#3{{\it Phys. Lett. }{\bf B#1} #3 (19#2)}

\end{document}